\documentclass{ws-procs961x669}            % default, citations in superscript
\begin{document}
\title{Different ways to estimate graviton mass¶
}
\author{Alexander F. Zakharov $^{1,2,3,4,5,*}$}

\address{$^{1}$National Astronomical Observatories of Chinese Academy of Sciences,
          20A Datun Road,  100012 Beijing, China\\
%\address{
$^{2}$Institute of Theoretical and Experimental
Physics,
            B. Cheremushkinskaya 25,
             117218 Moscow,  Russia \\
$^*$E-mail: zakharov@itep.ru\\
%}
%\address{
$^{3}$Bogoliubov
Laboratory of Theoretical Physics, Joint Institute
  for Nuclear Research,
   141980 Dubna,
             Russia\\
             %}
%\address{
$^{4}$National Research Nuclear University  MEPhI
(Moscow Engineering Physics Institute), 115409, Moscow,
 Russia\\
 % }
%\address{
$^{5}$North Carolina Central University, Durham, NC
27707,
 USA
 }

%\ead{zakharov@itep.ru}

\author{Predrag Jovanovi\'{c}$^6$}

\address{$^6$Astronomical Observatory, Volgina 7, 11060 Belgrade,
Serbia}

\author{Dusko Borka$^{7}$ and Vesna Borka Jovanovi\'{c}$^7$}

\address{Atomic Physics Laboratory (040), Vin\v{c}a
Institute of Nuclear Sciences \& \\ University of Belgrade, P.O. Box
522, 11001 Belgrade, Serbia}

\begin{abstract}
An experimental detection of graviton is extremely hard problem, however, there are different ways to evaluate a graviton
mass if it is non-vanishing.  Theories of massive gravity or theories with non-vanishing graviton mass initially have a number of pathologies such as discontinuities, ghosts etc. In last years theorists found ways to overcome weaknesses of such theories meanwhile observational features are also discussed. In the first publication reporting about the discovery of gravitational waves from the binary black hole system the LIGO-Virgo collaboration  obtained the graviton mass constraint around $1.2 \times 10^{-22}$~eV (later the estimate was improved with new data). A comparable and consistent graviton mass constraint around $2.9 \times 10^{-21}$~eV has been obtained from analysis of the bright star S2 trajectory near the Galactic Center.
\end{abstract}

\keywords{Gravitational waves; Black holes; Black hole physics; Graviton mass.}

\bodymatter

\section{Theories of massive gravity}

A few years ago Freeman Dyson discussed an opportunity to detect graviton and he concluded that at the moment an experimental detection of graviton is an extremely hard problem \cite{Dyson_13}, however, if a graviton has a mass as it is done in the framework of  theories of massive gravity, then there are different ways to constrain its mass. In the paper we discuss the issue in more details.
A  theory of massive gravity has been introduced by M. Fierz and W. Pauli in 1939 \cite{Fierz_39}.
In seventies a discontinuity of such approach for $m_g \rightarrow 0$ (where $m_g$ is a graviton mass) has been found \cite{Zakharov_70,vanDam_70,Iwasaki_70}. However, as it was   the so-called screening could resolve the issue\cite{Vainshtein_72} (see also subsequent studies in \cite{Rubakov_08,Babichev_09}) and the approach is very close to GR within natural assumptions.

Another pathology had been described\cite{Boulware_72,Boulware_72b} where the authors found a presence of ghosts and inconsistencies in gravity theories with a finite range, now such phenomena are usually called as Boulware -- Deser ghosts.
Different ways to create ghost-free massive gravity theories were discussed by Rubakov and Tinyakov\cite{Rubakov_08} including options when Lorentz invariance is violated.
Some years ago, two parameter family graviton potential or ghost-free massive gravity has been proposed \cite{deRham_10,deRham_11}, which is called now as de Rham -- Gabadadze -- Tolley (dRGT)
gravity model (see also more recent reviews \cite{deRham_14,deRham_17} for more comprehensive discussions).

A bi-metric theory of gravity with a massive graviton and its cosmological consequences were subjects of intensive studies by academician A. A. Logunov and his group \cite{Logunov_88,Chugreev_89,Gershtein_03,Gershtein_04,Logunov_06,Gershtein_07,Chugreev_15,Chugreev_16,Chugreev_17} at the period when such alternative theories of gravity  were not very popular as nowadays and a few decades ago it was a common opinion that a graviton mass must be vanishing due to presence of pathologies such as discontinuities  and ghosts.

Due to intensive developments in last years  theories of massive gravity may be treated now as respectable approaches and it would be reasonable to discuss their features and possible differences from conventional GR in observational predictions.

\section{Different observational constraints for graviton mass}

Perhaps first estimates of graviton mass have been given F. Zwicky \cite{Zwicky_61} as $m_g < 5 \times
10^{-64}$g since according to his opinion the Newtonian law has to be valid for galactic clusters\footnote{Similar idea had been used in paper
\cite{Hiida_65} where the authors obtained the estimate $m_g < 5 \times
10^{-62}$g (the authors also discussed also opportunity to constrain a graviton mass from Solar system data). As it was noted by V. L. Ginzburg, in astronomy ten is equal to one, but for so small (or big) quantities very often one hundred (or even one thousand) is equal to one, moreover, in astronomy estimates for lengths and masses (and related quantities) are significantly changing with time because an evaluation of lengths in astronomy is often model dependent.} (a few years earlier he discuss an opportunity that
the fundamental gravity law has to be valid for scales of galactic clusters because it is checked in a reliable way while for longer scales it may be changed \cite{Zwicky_57} and for such scales an impact of the cosmological $\Lambda$-term  may be significant).
M. Hare used Galactic scale to estimate Compton mass for graviton \cite{Hare_73}, while typical size of galactic clusters has been used  in paper \cite{Goldhaber_74}, thus the authors obtained  $m_g < 2 \times
10^{-62}$g (for a typical galactic cluster size  $\lambda_g < 3.7$~Mpc). If one uses the cosmological length scale \cite{Gershtein_03,Gershtein_04},
then $m_g < m_H^0$, where $m_H^0= \dfrac{\hbar H_0}{c^2}=3.8 h \times 10^{-66}$~g is "Hubble mass" and $H_0= h_{100} \times$ 100 km/(s Mpc) is the current Hubble constant ($h_{100}$ is a useful dimensionless parameter).

Practically, in first papers where people discussed observational constraints, there exist not real graviton mass estimates but expectations from future observations and related theoretical analysis because possible uncertainties and systematics were not taken into account. Therefore, many estimates are model dependent.

Observations of pulsars could give a nice opportunity to evaluate a graviton mass. Binary pulsars provide a remarkable test of GR predictions that
their orbits have to be shrinking due to gravitational radiation and it firstly was observed for Hulse -- Taylor pulsar PSR B1913+16.
In addition, it was shown\cite{Finn_02} evolutions of orbits for the binary pulsars PSR B1913+16 and PSR
B1534+12 constrained  the graviton mass at a level  $7.6 \times 10^{-20}$~eV with 90\% C.L.
As it was noted many years ago predicted and observed time of arrivals for pulsar residuals may be used for detection of gravitational waves.\cite{Sazhin_78} Current and future
pulsar timing arrays may be used not only for detection of gravitational waves but also to obtain  graviton mass constraints for different programs for observations at a level around
$[3 \times 10^{-23}, 3 \times 10^{-22}]$~eV as it was shown by Lee et al.\cite{Lee_10} and these constraints can be improved with a more sophisticated data analysis.\cite{Lee_13}
A number of other options to constrain a graviton mass from observations are given in reviews.\cite{deRham_17,Goldhaber_10}

\section{Graviton mass constraints from the discovery of gravitational waves}

\subsection{C. Will's idea}

Around 20 years ago C. Will considered an impact on observations of gravitational waves from the assumption that gravitons are massive \cite{Will_98} (see also \cite{Will_14}), then one has the well-known dispersion relation
\begin{equation}
E^2= \frac{m_g^2 c^4}{{1-(v_g^2/c^2)}}, \label{Dispersion_Eq}
\end{equation}
where $E$, $v_g$ and  $m_g$ are graviton energy, its velocity and mass, respectively. Therefore, to evaluate a graviton mass one has to compare a speed of gravity
and speed of light from observations. If there exist a source of gravitational waves and electromagnetic radiation such as supernovae or $\gamma$-ray burst (GRB)
than one can evaluate $1 - v_g/c$ \cite{Will_98}
\begin{equation}
1-\frac{v_g}{c}=5 \times 10^{-17}\left(\frac{200~\mathrm{Mpc}}{D} \right)\left(\frac{\Delta t}{1~\mathrm {s}} \right), \label{Dispersion_Eq 2}
\end{equation}
and
\begin{equation}
\Delta t = \Delta t_a - (1+z)\Delta t_e, \label{Dispersion_Eq 3}
\end{equation}
where $\Delta t_a = t^{EM}_a - t^{GW}_a$,  $\Delta t_e = t^{EM}_e - t^{GW}_e$, $t^{EM}_a (t^{EM}_e)$ and  $t^{GW}_a (t^{GW}_e)$ are
arrival (emission) time of electromagnetic radiation
and arrival (emission) time for gravitational waves. As it was noted \cite{Will_98}, usually $\Delta t_e$ is not known because there exist uncertainties in a source of electromagnetic radiation and  gravitational waves, however, sometimes one can evaluate an upper limit for the quantity.
Assuming that a graviton mass is small in comparison with energy of gravitational waves\cite{Will_98} $h f \gg m_g c^2$, then $v_g/c \approx 1- \dfrac{1}{2}\left(\dfrac{c}{\lambda_g f}\right)^2$, where $\lambda_g=h/(m_g c)$ is the graviton Compton wavelength and one can re-write
Eq. \eqref{Dispersion_Eq 2} as  it was done\cite{Will_98}
\begin{equation}
\lambda_g > 3 \times 10^{12} \mathrm{km} \left(\frac{D}{200~\mathrm{Mpc}} \frac{100~\mathrm {Hz}}{f}\right)^{1/2}\left(\frac{1}{f\Delta t} \right)^{1/2}, \label{Dispersion_Eq 4}
\end{equation}
As it was noted \cite{Will_98}, one can use  Eq. \eqref{Dispersion_Eq 4} if there exist gravitational and electromagnetic radiation from one source at a detectable level and $\Delta t_e$ is known or can be evaluated with a sufficient accuracy.
However, one can constrain  a graviton mass even in the case if there is only a gravitational wave signal without electromagnetic counterpart at a detectable level.\cite{Will_98} Really, if graviton is massive then graviton with higher frequencies propagate faster and as a result a gravitational wave signal will be different from a signal for a theory with massless graviton. For numerical estimate, one can change $f \Delta t \sim \rho^{-1} \approx 10$ (where $\rho$ is a signal-to-noise ratio) for LIGO-Virgo ground based interferometers.\cite{Will_98} For LISA space-based detectors a graviton mass constraint can be much smaller, so it may be at a level $2.5 \times 10^{-22}$~eV for ground based LIGO-Virgo detectors and  at a level $2.5 \times 10^{-26}$~eV for LISA.\cite{Will_98} A combined observations of 50 events in two-year LISA mission can improve previous estimates of graviton mass.\cite{Berti_11}

\subsection{Graviton mass constraints from first LIGO events}
In February 2016 the LIGO-Virgo collaboration reported about the first detection of gravitational waves from a merger of two black holes (it was detected on September 15, 2015 and it is called GW150914)\cite{Abbott_16}. The source is located at a luminosity distance of around 410~Mpc (which corresponds to a redshift $z \approx 0.09$. The initial black hole masses were $36 M_\odot$ and
$29 M_\odot$  and the final black hole mass is
$62 M_\odot$, therefore around   $3 M_\odot$ radiated in gravitational waves in 0.1 s. The collaboration not only discover gravitational waves but also found the first binary black hole system and one of the most powerful source of radiation in the Universe and energy was release in gravitational waves. Moreover, the team constrained the graviton Compton wavelength $\lambda_g > 10^{13}$~km which could be interpreted as a constraint a graviton mass
$m_g < 1.2 \times 10^{-22}$~eV.\cite{Abbott_16} The authors analyzed possible changes of the dispersion relation for massive gravitons as it was discussed in paper.\cite{Will_98}

In June 2017 the LIGO-Virgo collaboration published paper where the authors described a detection of gravitational wave signal from a merger of binary black hole system with masses of components $31.2M_\odot$ and $19.4 M_\odot$ at distance around 880~Mpc which corresponds to $z \approx 0.18$.\cite{Abbott_17} In this case, around $2M_\odot$ were emitted in gravitational waves in around 0.4~s. The event was observed on January 4, 2017 and it is named GW170104. In this paper the authors improved their previous graviton mass constraint almost in two times, $m_g  < 7.7 \times 10^{-23}$~eV.\cite{Abbott_17}

\subsection{Graviton mass constraint and the neutron star merger GW170817}
On August 17, 2017  the LIGO-Virgo collaboration observed a merger of binary neutron stars with masses around $0.86 M_\odot$ and $2.26 M_\odot$ at a distance around 40 Mpc (GW170817) and after 1.7 s the Fermi-GBM found $\gamma$-ray burst GRB 170817A associated with the GW170817.\cite{Abbott_170817,Abbott_ApJL_170817}  Since gravitational wave signal was detected before GRB one could conclude that
the observational data are consistent with massless or very light graviton, otherwise, electromagnetic signal could be detected before gravitational one because in the case of relatively heavy gravitons gravitational waves could propagate slower than light. Constraints on speed of gravitational waves have been found\cite{Abbott_ApJL_170817} $-3\times 10^{-15} < (v_{g}-c)/c < 7 \times 10^{-16}$. Graviton energy is $E=hf$, therefore, assuming a typical LIGO frequency range $f \approx 100$, from Eq. \eqref{Dispersion_Eq} one could obtain a graviton mass estimate $m_g < 3.2 \times 10^{-20}$~eV which a slightly weaker estimate than previous ones obtained from binary black hole signals detected by the LIGO team.

\section{Graviton mass constraints from analysis of trajectories of bright stars at the Galactic Center}

Observations of bright stars in IR band provide a very efficient tool to evaluate a gravitational potential at the Galactic Center (see, papers \cite{Gillesen_17,Hees_17} and references therein). Such observations give an opportunity to evaluate parameters of black hole and bulk distribution of mass in a stellar cluster and dark matter \cite{Zakharov_07}. One could use these data to constrain alternative theories of gravity such as
$R^n$ \cite{Borka_12} or Yukawa theory. \cite{Borka_13} If we apply our consideration for gravity theories with massive graviton and we use observational data for S2 star we obtain that
$2.9 \times 10^{-21}$~eV with 90\% C.L.\cite{Zakharov_JCAP_16} (see also discussion\cite{Zakharov_Quarks_16,Zakharov_Baldin_17,Zakharov_MIFI_17}).

%\section{Conclusions}
Our estimate for graviton mass is slightly weaker than the LIGO ones, but it is independent and consistent with LIGO results. In the future the current graviton mass estimate obtained from analysis of S2 data can be significantly  improved with forthcoming facilities such GRAVITY, TMT and E-ELT.

\section*{Acknowledgments}
A. F. Z. thanks
PIFI grant 2017VMA0014 of Chinese Academy of Sciences, the Strategic Priority Research Program of the Chinese Academy of Sciences (Grant No. XDB23040100),
NSF
(HRD-0833184) and NASA (NNX09AV07A) at NASA CADRE and NSF CREST
Centers (NCCU, Durham, NC, USA) for a partial support.
A. F. Z. also thanks the organizers of the XXXI International Workshop on High Energy Physics (IHEP, Protvino)
for their attention to our contribution. P.J., D.B. and V.B.J. wish to acknowledge the support by the Ministry
of Education, Science and Technological Development of the Republic of
Serbia through the project 176003 ''Gravitation and the large scale structure
of the Universe''.

%Non BiBTeX users can list down their references as:


\begin{thebibliography}{100}

\bibitem{Dyson_13}
F. Dyson,
{\em Intern. J. Mod. Phys.} A {\bf 28}, 1330041 (14 pages) (2013).


\bibitem{Fierz_39}
M. Fierz and W. Pauli,
{\em Proc. of the Royal Society of London} {\bf A173}, 211 (1939).


\bibitem{Zakharov_70}
V. I. Zakharov,  {\em JETP Lett.} {\bf 12}, 447 (1970).

\bibitem{vanDam_70}
H. van Dam and M.
Veltman, {\em Nucl. Phys.} B {\bf 22}, 397 (1970).

\bibitem{Iwasaki_70}
Y. Iwasaki, {\em Phys. Rev.} D
{\bf 2}, 2255 (1970).

\bibitem{Vainshtein_72}
A.I. Vainshtein,  {\em Phys. Lett.} B {\bf 39}, 393 (1972).
%
\bibitem[\protect\citeauthoryear{Rubakov and Tinyakov}{2008}]{Rubakov_08}
V. A. Rubakov and P. G. Tinyakov,  {\em Physics -- Uspekhi} {\bf 51}, 759
(2008).


\bibitem{Babichev_09}
E. Babichev,  C. Deffayet and R. Ziour,
%Recovering General Relativity from massive gravity,
{\em Phys. Rev. Lett.}
{\bf 103}, 201102 (2009).
% arXiv:0907.4103 [gr-qc].

\bibitem{Boulware_72}
D. G. Boulware and S. Deser,
%Can Gravitation Have a Finite Range?
{\em Phys. Rev.}
{\bf 6},  3368 (1972).

\bibitem{Boulware_72b}
D. G. Boulware and S. Deser,
%Inconsistency of finite range gravitation
{\em Phys. Lett.} B
{\bf 40},  227 (1972).



\bibitem{deRham_10}
C. de Rham and G. Gabadadze,
%Generalization of the Fierz-Pauli Action,
{\em Phys.Rev.} D {\bf 82},
044020 (2010).
% arXiv:1007.0443 [hep-th]

\bibitem{deRham_11}
C. de Rham,  G. Gabadadze and A. J.
Tolley,
%Resummation of Massive Gravity,
{\em Phys. Rev. Lett.} {\bf 106}, 231101 (2011).

\bibitem{deRham_14}
C. de Rham,
%\Massive Gravity,
{\em Living Rev.
Rel.} {\bf 17}, 7 (2014).
% arXiv:1401.4173 [hep-th]

\bibitem%[\protect\citeauthoryear{de Rham et al.}{2017}]
{deRham_17}
C. de Rham, J. T. Deskins, A. J. Tolley
{\it et al.},
%and  S.-Y. Zhou,
{\em Rev. Mod. Phys.} {\bf  89}, 025004 (2017).

\bibitem{Visser_98}
M. Visser,  {\em Gen. Rel. Grav.}  {\bf 30}, 1717 (1998).




\bibitem{Logunov_88}
A. A. Logunov, M. A. Mestvirishvili
and Yu. V. Chugreev,
%Graviton mass and evolution of a Friedmann universe
{\em Theor.  Math. Phys.} {\bf 74},
  1 (1988).

\bibitem{Chugreev_89}
Yu. V. Chugreev,
%Cosmological consequences of the relativistic theory of gravitation with massive gravitons
{\em Theor.  Math. Phys.} {\bf 79},
  554 (1989).


\bibitem{Gershtein_03}
S. S. Gershtein,  A. A. Logunov and M. A. Mestvirishvili,
%Graviton Mass and the Total Relative Mass Density $\Omega_tot$
%in the Universe
{\em Dokl. Phys.} {\bf 48} 282  (2003).


\bibitem{Gershtein_04}
S. S. Gershtein, A. A. Logunov,  M. A. Mestvirishvili and N. P. Tkachenko,
%Graviton Mass, Quintessence, and Oscillatory Character of Universe
%Evolution,
{\em Phys. Atom. Nucl.} {\bf 67},
  1596 (2004).

\bibitem{Logunov_06}
A. A. Logunov,
{\em Relativistic Theory of Gravitation},
(Nauka, Moscow, 2006, in Russian).


\bibitem{Gershtein_07}
S. S. Gershtein, A. A. Logunov and M. A. Mestvirishvili,
%Cosmological Constant and Minkowski Space
{\em Phys. Part.  Nucl.} {\bf 38}, 291 (2007).

\bibitem{Chugreev_15}
Yu. V. Chugreev,
%Mach's Principle for Cosmological Solutions
%in Relativistic Theory of Gravity
{\em Phys. Part.  Nucl. Lett.} {\bf 12}, 195 (2015).

\bibitem{Chugreev_16}
Yu. V. Chugreev,
%Dark Energy and Graviton Mass in the Nearby Universe,
{\em Phys. Part.  Nucl. Lett.} {\bf 13}, 38 (2016).

\bibitem{Chugreev_17}
Yu. V. Chugreev,
%Cosmological Constraints on the Graviton Mass in RTG
{\em Phys. Part.  Nucl. Lett.}  {\bf 14}, 539 (2017).

\bibitem{Zwicky_61}
F. Zwicky, {\em Publ. Astron. Soc. Pac.} {\bf 73},
314 (1961).

\bibitem{Hiida_65}
K. Hiida and Y. Yamaguchi,
%Gravitation Physics
 {\em Prog. Theor. Phys. Suppl.} E {\bf 65},
261 (1965).


\bibitem{Zwicky_57}
F. Zwicky, {\em Publ. Astron. Soc. Pac.} {\bf 69},
518 (1957).

\bibitem{Hare_73}
M. G. Hare,
{\em Can. J. Phys.}  {\bf 51}, 431 (1973).

\bibitem{Goldhaber_74}
A. S. Goldhaber and M. M. Nieto,
{\em Phys. Rev.} D {\bf 9}, 1119 (1974).



\bibitem[\protect\citeauthoryear{Finn \& Sutton}{2002}]{Finn_02}
L.S. Finn and P.J. Sutton, {\em Phys. Rev.} D {\bf 65}, 044022 (2002).

\bibitem{Sazhin_78}
M. V. Sazhin, {\em Sov. Astron.} {\bf  22}, 36  (1978).

\bibitem{Lee_10}
K. Lee, F. A. Jenet,  R. H. Price {\it et al.},
{\em Astrophys. J.}  {\bf 722}, 1589 (2010).


\bibitem{Lee_13}
K. Lee, %Pulsar timing arrays and gravity tests in the radiative regime,
{\em Class. Quant. Grav.}  {\bf 30} (2013) 224016 (12pp).

\bibitem{Goldhaber_10}
A. S. Goldhaber and M. M. Nieto,
{\em Rev. Mod. Phys. }  {\bf 82}, 939 (2010).


\bibitem{Will_98}
C. Will,
%Bounding the mass of the graviton using gravitational-wave observations
%of inspiralling compact binaries
{\em Phys. Rev.} D {\bf 57}, 2061 (1998).

\bibitem{Will_14}
C. Will,
%\emph{The Confrontation between General Relativity and
%Experiment},
{\em Liv. Rev. Relat.}  {\bf 17},  4 (2014).

\bibitem{Berti_11}
E. Berti, J. Gair and A. Sesana,
%Graviton mass bounds from space-based gravitational-wave observations
%of massive black hole populations
{\em Phys. Rev.} D {\bf 84}, 101501(R) (2011).

\bibitem%[\protect\citeauthoryear{Abbott et al.}{2016}]
{Abbott_16}
B. P. Abbott {\it et al.},
%(LIGO Scientific Collaboration and Virgo
%Collaboration)
%2016
%Observation of Gravitational Waves from a
%Binary Black Hole Merger,
{\em Phys. Rev. Lett.}  {\bf 116}, 061102 (2016).

\bibitem%[\protect\citeauthoryear{Abbott et al.}{2016}]
{Abbott_17}
B. P. Abbott {\it et al.},
%(LIGO Scientific Collaboration and Virgo
%Collaboration)
%2016
%GW170104: Observation of a 50-Solar-Mass Binary Black Hole Coalescence
%at Redshift 0.2,
{\em Phys. Rev. Lett.}  {\bf 118}, 221101 (2017).


\bibitem%[\protect\citeauthoryear{Abbott et al.}{2016}]
{Abbott_170817}
B. P. Abbott {\it et al.},
%GW170817: Observation of Gravitational Waves from a Binary Neutron Star Inspiral
{\em Phys. Rev. Lett.}  {\bf 119}, 161101 (2017).

\bibitem%[\protect\citeauthoryear{Abbott et al.}{2016}]
{Abbott_ApJL_170817}
B. P. Abbott {\it et al.},
%Gravitational Waves and Gamma-Rays from a Binary Neutron Star Merger:
%GW170817 and GRB 170817A
{\em Astrophys. J.  Lett.}  {\bf 848}, L13 (27pp) (2017).

\bibitem%[Gillessen et al.(2017)]
{Gillesen_17}
S. Gillessen, P.~M. Plewa, F. Eisenhauer {\it et al.}, {\em Astrophys. J.}
{\bf 837}, 30 (2017).

\bibitem[\protect\citeauthoryear{Hees et al.}{2017}]{Hees_17}
A. Hees, T. Do, A. M. Ghez {\it et al.},
%Testing General Relativity with stellar orbits around the supermassive black hole in our Galactic center,    arXiv:1705.07902v1 [astro-ph.GA],
{\em Phys. Rev. Lett.} {\bf 118}, 211101 (2017).
%, arXiv:1705.07902v1 [astro-ph.GA]


\bibitem{Zakharov_07}
A.F. Zakharov, A.A.  Nucita, F.  De Paolis
{\it et al.},
%and G.  Ingrosso,
{\em Phys.
Rev. D} {\bf 76}, 062001 (2007).

\bibitem[\protect\citeauthoryear{Borka et al.}{2012}]{Borka_12}
D. Borka, P.  Jovanovi\'c, V. Borka Jovanovi\'c
{\it et al.},
%and  A.F.  Zakharov,
{\it Phys. Rev.} D {\bf 85}, 124004 (2012).

\bibitem[\protect\citeauthoryear{Borka et al.}{2013}]{Borka_13}
D. Borka
%, P.  Jovanovi\'c, V. Borka Jovanovi\'c
{\it et al.},
%and  A.F. Zakharov,
{\it J. Cosm. Astropart. Phys.} {\bf 11}, 050 (2013).

\bibitem{Zakharov_JCAP_16}
A. F. Zakharov {\it et al.},
%, Jovanovi{\'c} P, Borka D,  Borka Jovanovi{\'c} V
%2016
{\em J. Cosmol. Astropart. Phys.} JCAP {\bf 05},   045 (2016).

\bibitem{Zakharov_Quarks_16}
A. F. Zakharov  {\it et al.},
%, Jovanovi{\'c} P, Borka D,  Borka Jovanovi{\'c} V
{\em EPJ Web of Conferences} {\bf 125}, 01011 (2016).

\bibitem{Zakharov_Baldin_17}
A. F. Zakharov  {\it et al.},
%P. Jovanovi{\'c}, D. Borka, V. Borka Jovanovi{\'c},
%Graviton mass bounds from an analysis of bright star trajectories at the Galactic Center,
%in Proceedings of the XXIII International Baldin Seminar on High Energy Physics Problems Relativistic Nuclear Physics and Quantum Chromodynamics (Baldin ISHEPP XXIII), eds. S. Bondarenko, V. Burov and A. Malakhov
{\em EPJ Web of Conferences} {\bf 138}, 01010 (2017).

\bibitem{Zakharov_MIFI_17}
A. F. Zakharov  {\it et al.},
%P. Jovanovi{\'c}, D. Borka, V. Borka Jovanovi{\'c},
%Graviton mass evaluation with trajectories of bright stars at the Galactic Center,
%in Proceedings of the International Conference on Particle Physics and Astrophysics
%10–14 October 2016, Moscow, Russia, eds. A M Galper, A A Petrukhin, S G Rubin, I V Selyuzhenkov, M D Skorokhvatov, E Soldatov and S A Voronov,
{\em
Journal of Physics: Conference Series} {\bf 798}, 012081 (2017).













%\bibitem[\protect\citeauthoryear{Dvali et al.}{2000}]{Dvali_00}
%G. Dvali, G. Gabadadze and M. Porrati, {\em Phys. Lett.} B {\bf 485}, 208
%(2000).
%
%\bibitem[\protect\citeauthoryear{Kogan et al.}{2001}]{Kogan_01}
%I. I. Kogan, S. Mouslopoulos, A. Papazoglou and G. G. Ross, Nucl. Phys.
%B {\bf 595},  225 (2001).
%
%\bibitem[\protect\citeauthoryear{Deffayet et
%al.}{2002}]{Deffayet_02}
%C. Deffayet G. Dvali, G. Gabadadze
%and M. Porrati, Phys. Rev. D {\bf 65}, 044026 (2002).



%\bibitem[\protect\citeauthoryear{Damour, Kogan \& Papazoglou}{2003}]{Damour_03}
%T. Damour, I.I. Kogan and A. Papazoglou, Phys. Rev. D {\bf 67},
%064009 (2003).









%\bibitem{lamp94}
%L.~Lamport, {\em \LaTeX, A Document Preparation System}, 2nd edn.
%  (Addison-Wesley, Reading, MA, 1994).
%
%\bibitem{ams04}
%\AmS, {\em \AmS-\LaTeX{} Version 2 User's Guide} (American Mathematical
%  Society, Providence, 2004),
%  \url{http://www.ams.org/tex/amslatex.html}.
%
%\bibitem{jarl88}
%C.~Jarlskog, {\em CP {V}iolation} (World Scientific, Singapore, 1988).
%
%\bibitem{best03}
%B.~W. Bestbury, $R$-matrices and the magic square, {\em J. Phys. A} {\bf 36},
%  1947 (2003).
%
%\bibitem{pier02}
%P.~X. Deligne and B.~H. Gross, On the exceptional series, and its descendants,
%  {\em C. R. Math. Acad. Sci. Paris} {\bf 335}, 877  (2002).
%
%\bibitem{jame02}
%J.~M. Landsberg and L.~Manivel, Triality, exceptional Lie algebras and
%  Deligne dimension formulas, {\em Adv. Math.} {\bf 171}, 59  (2002),
%  \url{http://www.url.com/triality.html}.
%
%\bibitem{weis94}
%G.~H. Weiss (ed.), {\em Contemporary {P}roblems in {S}tatistical {P}hysics}
%  (SIAM, Philadelphia, 1994).
%
%\bibitem{gupt97}
%R.~K. Gupta and S.~D. Senturia, Pull-in time dynamics as a measure of absolute
%  pressure, in {\em Proc. IEEE Int. Workshop on Microelectromechanical
%  Systems ({MEMS}'97)\/},  (Nagoya, Japan, 1997).
%
%\bibitem{rich60}
%L.~F. Richardson, {\em Arms and Insecurity} (Boxwood, Pittsburg, 1960).
%
%\bibitem{chur90}
%R.~V. Churchill and J.~W. Brown, {\em Complex Variables and Applications},
%  5th edn. (McGraw-Hill, 1990).
%
%\bibitem{benh93}
%F.~Benhamou and A.~Colmerauer (eds.), {\em Constraint Logic Programming,
%  {S}elected {R}esearch} (MIT Press, 1993).
%
%\bibitem{bake72}
%D.~W. Baker and N.~L. Carter, {\em Seismic {V}elocity {A}nisotropy {C}alculated
%  for {U}ltramafic {M}inerals and {A}ggregates}, in {\em Flow and {F}racture of
%  {R}ocks\/},  eds. H.~C. Heard, I.~V. Borg, N.~L. Carter and C.~B. Raleigh,
%  Geophys. Mono., Vol.~16 (Am. Geophys. Union, 1972), pp. 157--166.
%
%\bibitem{hobb92}
%J.~D. Hobby, {\em {A User's Manual for MetaPost}}, Tech. Rep. 162, AT\&T Bell
%  Laboratories (Murray Hill, New Jersey, 1992).
%
%\bibitem{bria84}
%B.~W. Kernighan, {\em {PIC}---{A} Graphics Language for Typesetting}, Computing
%  Science Technical Report 116, AT\&T Bell Laboratories (Murray Hill, New
%  Jersey, 1984).
%
%\bibitem{hear94}
%H.~C. Heard, I.~V. Borg, N.~L. Carter and C.~B. Raleigh, {VoQS: Voice Quality
%  Symbols}, Revised to 1994,  (1994).
%
%\bibitem{brow88}
%M.~E. Brown, An interactive environment for literate programming, PhD thesis,
%  Texas A\&M University, (TX, USA, 1988), pp. ix + 102.
%
%\bibitem{lodh74}
%G.~S. Lodha, Quantitative interpretation of ariborne electromagnetic response
%  for a spherical model, Master's thesis, University of Toronto  (1974).
%
%\bibitem{dani73}
%D.~Jones, {The term `phoneme'}, in {\em Phonetics in Linguistics: A Book of
%  Reading\/},  eds. W.~E. Jones and J.~Laver (Longman, London, 1973) pp.
%  187--204.
%
%\bibitem{davi93}
%B.~Davidsen, Netpbm  (1993),
%  \url{ftp://ftp.wustl.edu/graphics/graphics/packages/NetPBM}.

\end{thebibliography}
\end{document}